\documentclass[twocolumn,showpacs,amsmath]{revtex4}

\usepackage[T1]{fontenc}
\usepackage[cp1250]{inputenc}
\usepackage{graphicx}
\usepackage{dcolumn}
\usepackage{bm}
\usepackage{ulem}

\newcommand{\vect}[1]{{\boldsymbol{\mathbf{#1}}}}

\begin{document}

\title{Electromagnetically Induced Transparency versus Nonlinear Faraday
Effect.

Coherent Control of the Light Beam Polarization}
\author{R.~Drampyan}
\affiliation{Institute for Physical Research, National Academy of Sciences of Armenia, \\0203 Ashtarak-2, Armenia}
\author{S.~Pustelny, W.~Gawlik}
\affiliation{Center for Magneto-Optical Research,
Institute of Physics, Jagiellonian University, Reymonta 4, 30-059
Krak\'ow, Poland}

\date{\today}

\begin{abstract}
We report on experimental and theoretical study of the nonlinear
Faraday effect under conditions of electromagnetically induced
transparency at the 5$S_{1/2} \rightarrow 5P_{3/2} \rightarrow 5D_{5/2}$ two-photon
transition in rubidium vapors. These transitions realize the
inverted Y model which combines the $\Lambda$ and ladder systems.
Strong nonlinearity allowing for large rotation angles of a probe
beam tuned to the $S\rightarrow P$ transition was obtained by
creation of quantum superpositions of magnetic sublevels (Zeeman
coherences) in the rubidium ground state ($\Lambda$ scheme).
Additionally, electromagnetically induced transparency was
accomplished in a ladder scheme by acting with an additional
strong coupling laser on the $P\rightarrow D$ transition. Under
conditions of a two-photon resonance the rotation was
significantly reduced, which is interpreted as a competition
between the two processes. The effect was observed in sub-Gauss
magnetic fields and could be used for efficient coherent control
of generation of the ground-state coherences,
e.g. for controlling the polarization state of the probe
light.
\end{abstract}

 \maketitle

\section{Introduction\label{sec:Introduction}}

Three-level atomic systems allow observation of several
interesting and widely studied quantum-interference phenomena,
such as coherent population trapping \cite{Arimondo1996},
electromagnetically induced transparency \cite{Fleischhauer2005},
electromagnetically induced absorption \cite{Akulshin1998}.
Extension of a number of levels and/or fields results in new
effects related to double-dressing \cite{Lukin1999,Wasik2001}, and
interplay of coherences in the $N$ \cite{Echaniz2001,Goren2004},
tripod \cite{Kis2002,Petrosyan2004}, or inverted Y
\cite{Joshi2003,Joshi2005} structures.

A specific example of quantum interference phenomenon is nonlinear magneto-optical rotation \cite{NMORreview}. The effect is based on light-intensity dependent rotation of the polarization plane of linearly polarized light during its propagation through a medium subjected to the magnetic field. In this paper we address a special case of the phenomenon, the nonlinear Faraday effect (NFE) \cite{Gawlik2009}, which occurs in a magnetic field longitudinal to light propagation direction.

The rotation is associated with light-generated coherences between
magnetic sublevels of atomic or molecular ground and/or excited
states. NFE has thus the same physical mechanism as coherent
population trapping (CPT) and the advantage of better
signal-to-noise ratio over transmission measurements performed in
most of the CPT experiments. A typical NFE signal recorded versus
magnetic field is characterized by a dispersive shape centered at
zero field with the width and amplitude determined by the
relaxation rate of the coherences. Particularly large rotations
(up to 100~mrad) can be observed when the coherences
are established between magnetic sublevels of long-lived ground
state. Application of anti-relaxation coating of gas container
inner walls or introduction of a buffer gas into the vapor cell
allow observation of extremely narrow, i.e. sub-$\mu$G, NFE
signals associated with coherences of 100-ms or longer lifetimes
\cite{NMORreview,Budker2005}. Another quantum interference
phenomenon relevant for our research is electromagnetically
induced transparency (EIT). The effect consists in reduction of
the absorption of a weak, resonant probe light propagating through
a medium in which the probe and a strong coupling beam jointly
establish two-photon coherence in the three-level system. EIT may
be observed in several three-level structures: $\Lambda$
\cite{Li1995,Hau1999}, V \cite{Boon1998} and ladder
\cite{Moseley1995,Clarke2001}. EIT, as well as CPT
are often used for inducing narrow transparency windows in
otherwise optically dense samples and/or generating steep
dispersion responsible for drastic modifications of the speed of
light \cite{Hau1999,Merriam1999,Budker1999}.

In this paper we analyze coexistence and competition between two
different effect related with atomic coherences: NFE and EIT.
Despite the fact that NFE is associated with ground-state
coherences and EIT with coherences in a ladder scheme, they share
intermediate excited levels and effectively affect one another.
Interplay between the phenomena makes it possible to perform
efficient all-optical coherent control over the polarization
rotation. The NFE signals are analyzed for different probe and
coupling-beam parameters and magnetic field strengths.
Experimental data is supported by a theoretical model
qualitatively explaining the observed characteristics.

Our approach significantly differs from the previous experiments
on coherent control of optical rotation
\cite{Liao1976,Liao1977,Pavone1997,Wielandy1998,Padney2008}. In particular,
in Refs. \cite{Liao1976,Liao1977,Pavone1997} no magnetic field was used and
the rotation arose exclusively due to light-induced birefringence
of the medium. Wielandy \textit{et al.} \cite{Wielandy1998} and
Padney \textit{et al.} \cite{Padney2008} did apply magnetic field
but it was much stronger ($\sim 10$~G) than in our work, therefore
the NFE observed in Refs.~\cite{Wielandy1998,Padney2008}
was due to the excited- rather than ground-state coherences.
Below, we demonstrate that involving the ground-state coherences
results in an effective control by sub-Gauss magnetic fields.

The paper is organized as follows. In Sec.~\ref{sec:Experiment}
the experimental arrangement is described. Experimental results
are discussed in Sec~\ref{sec:Results}. Section~\ref{sec:Theory}
presents theoretical interpretation of the observed effects which
are summarized in Sec.~\ref{sec:Conclusions}.

\section{Experimental apparatus\label{sec:Experiment}}

\begin{figure}
 \includegraphics[width=\columnwidth]{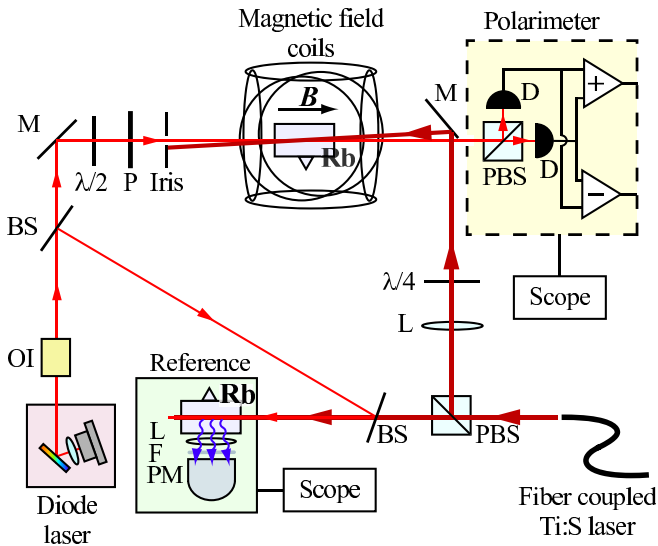}
 \caption{Experimental apparatus. Two counter-propagating laser beams
 tuned to $5S_{1/2}\rightarrow 5P_{3/2}$ (probe light) and $5P_{3/2}\rightarrow 5D_{5/2}$ (coupling light) transition interact with rubidium atoms subjected to longitudinal magnetic field. The probe light is generated by an external-cavity diode laser, while the coupling light is provided by fiber-coupled titanium sapphire laser (Ti:S). OI denotes optical isolator, BS beam splitter, PBS polarizing beam splitter, M mirror, L lens, F interference filter, $\lambda/2$ and $\lambda/4$ are half-wave and quarter-wave plates, respectively, Rb is the rubidium vapor cell, PM photomultiplier, and D photodiodes.}
 \label{fig:setup}
\end{figure}

\begin{figure}[htb!]
 \begin{center}
  \includegraphics{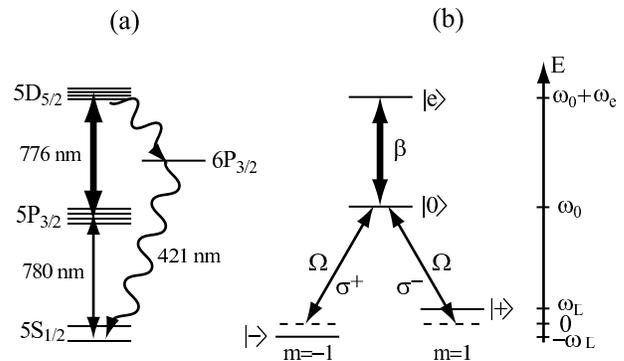}
 \end{center}
 \caption{Rubidium energy level structure (a) and inverted Y model used for theoretical
 analysis (b). Linearly polarized probe light was tuned to the $5S_{1/2}\rightarrow
 5P_{3/2}$ transition while the strong coupling beam excited atoms at the
 $5P_{3/2}\rightarrow 5D_{5/2}$ transition. The atoms deexcited from the $5D_{5/2}$
 state to the $5S_{1/2}$ state via intermediate state $6P_{3/2}$. Decay of the $6P_{3/2}$
 state was accompanied with emission of blue light (421 nm) which was monitored in our
 experiment and used as reference.}
 \label{fig:EnergyStructure}
\end{figure}

Experimental arrangement is shown in Fig.~\ref{fig:setup}.
Two counter-propagating beams: weak probe and strong coupling beam
of 1 and 3 mm diameters, respectively, were overlapped in a
rubidium vapor cell. The 4-cm long, 2.5-cm diameter cell
contained natural mixture of rubidium with neither buffer gas nor
wall coating. The probe light emitted from the external-cavity
diode laser ($\sim\!\!1$~MHz linewidth) was tuned to a particular
hyperfine transition of the rubidium $D2$ line
($5S_{1/2}\rightarrow 5P_{3/2}$ transition of 780~nm wavelength
and 5.8~MHz natural linewidth) or scanned across the whole
Doppler-broadened D2 line. The probe beam was linearly polarized
and its power was between 75 and 750~$\mu$W. The coupling beam
emitted from a CW titanium sapphire laser ($\sim\!3$~MHz
linewidth) was tuned to the $5P_{3/2}\rightarrow 5D_{5/2}$
transition corresponding to 776~nm wavelength and 0.97~MHz natural
linewidth and its power was varied up to 30~mW. The corresponding energy level
structure is shown in Fig.~\ref{fig:EnergyStructure}.

Polarization of the coupling beam was set either linear or circular by a
$\lambda$/4 waveplate. The counter-propagating arrangement of the
probe and coupling beams provided that the two-photon interaction
between the $5S_{1/2}$ and $5D_{5/2}$ states was possible only for
a specific atomic velocity class which allowed observation of
Doppler-free spectral features. The second rubidium cell was
illuminated with co-propagating fractions of the probe and coupling
beams and equipped with an interference filter and
photomultiplier. It served as a reference for the two-photon
resonance by monitoring the 421~nm fluorescence associated with
deexcitation of the 5D levels in the $5D_{5/2}\rightarrow
6P_{1/2}\rightarrow 5S_{1/2}$ cascade (Fig.~\ref{fig:EnergyStructure}). The main rubidium cell was kept at room
temperature (20$^\circ$C) that corresponded to atomic density of
$\sim\!10^{10}$~cm$^{-3}$ and a Doppler width of $\sim 0.5$~GHz.
The cell was surrounded by three orthogonal pairs of Helmholtz
coils which compensated stray magnetic fields. The same coils were
used for generation of an additional bias field in the range of
-3~G to 3~G along the laser-beam direction.

The probe beam rotation was measured with a balanced polarimeter,
i.e., a high-quality crystal polarizer and two photodiodes
detecting intensities of two beams emerging from the polarizer.
Analysis of the signals from two photodiodes yielded the angle of
polarization rotation and transmitted signal power. In the
described experiment, the rotation signals were recorded as
functions of a longitudinal magnetic field $B$ at fixed
laser frequencies, or as functions of the probe frequency
for fixed tuning of the coupling laser and constant magnetic field
(usually about 70~mG).

\section{Results\label{sec:Results}}

\subsection{NFE without coupling beam}

\begin{figure}
 \includegraphics[width=\columnwidth]{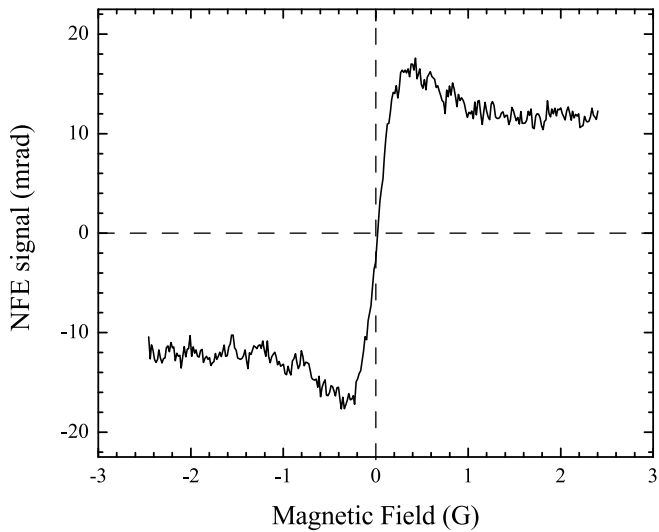}
 \caption{NFE signal measured vs. longitudinal magnetic field.
The signal was measured with the probe beam power of 75~$\mu$W
tuned to the center of the $F = 3\rightarrow F'$ component of the
$D2$ line in $^{85}$Rb, and no coupling beam.}
 \label{fig:NFE}
\end{figure}

The measurements started with recording of the probe-light
polarization rotation versus the longitudinal magnetic field
in absence of the coupling light. A typical signal
measured under such conditions is shown in Fig.~\ref{fig:NFE}.
It is characterized by an asymmetric curve of 20-mrad amplitude
and a width of $\sim 400$~mG around $B=0$. In a buffer-gas-free
uncoated vapor cell, such as used in this work, the ground-state
relaxation rate and hence the width of NFE resonance at
low laser power equals to the inverse transit-time of atoms across
the light beam. Under our experimental conditions the transit
relaxation time was on the order of a few $\mu$s which roughly
corresponds to a width of 200~mG. The difference between this
number and experimentally observed width is attributed to power
broadening caused by the 75 $\mu$W probe light.
\begin{figure}
 \includegraphics[width=\columnwidth]{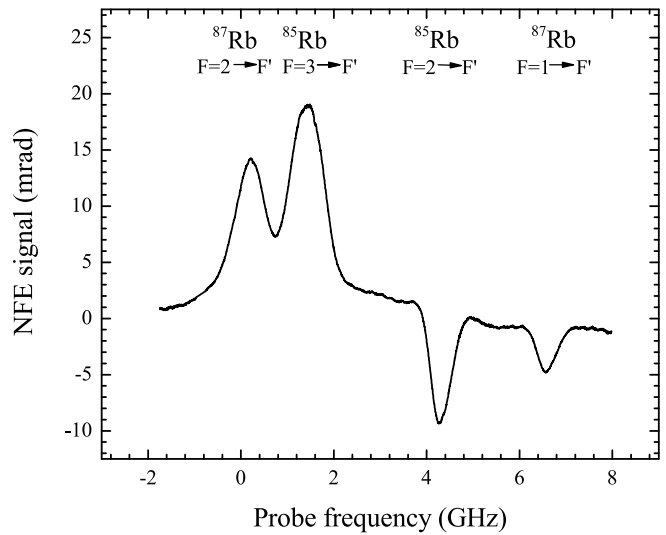}
 \caption{NFE spectrum recorded at $B=67$~mG without the coupling beam. The probe
 laser of 750~$\mu$W was scanned around the hyperfine components with F=2 and F=3
 of the D2 line of $^{87}$Rb and $^{85}$Rb, respectively.}
  \label{fig:SpectraWithoutCoupling}
\end{figure}
The resonance presented in Fig.~\ref{fig:NFE} has a shape which
results from superposition of several nested dispersive
contributions associated with Zeeman coherences in the ground and
excited states, which are all centered at $B=0$ but have different
widths. Without the coupling beam the rotation spectra,
i.e. NFE signal recorded versus the probe beam frequency $\omega_p$
at non-zero magnetic field ($B\neq0$), are Doppler-broadened curves centered at hyperfine and isotope components of the investigated line
(Fig.~\ref{fig:SpectraWithoutCoupling}). For
light power of 750~$\mu$W the rotation reaches 20~mrad already at
$B=67$ mG which illustrates strong magnetic-field dependence of
NFE, and maximizes at about 0.5 G at a level of a few tens of mrad.
In such a case, however, the NFE signals could be contaminated by
excited-state coherence (see discussion in Sec.~\ref{sec:Theory}),
so the measurements were performed for $B<100$~mG.

\subsection{NFE with a coupling beam}

\begin{figure}[h!]
 \includegraphics[width=\columnwidth]{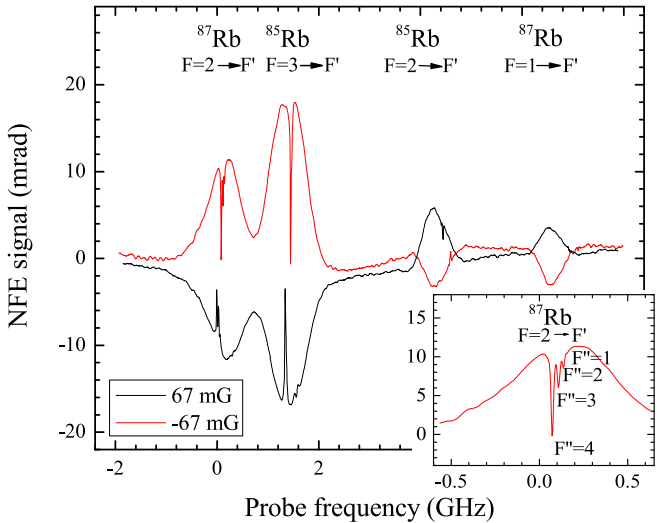}
 \caption{Typical NFE spectra obtained with fixed frequency of the coupling laser
 and the probe light scanned across the rubidium $D2$ line. The spectra consist of
 Doppler-broadened resonances and narrow spectral features. Inset shows the zoomed
 $F=2\rightarrow F'$ transition of $^{87}$Rb and reveals dips corresponding to
 excited-state hyperfine structure. The spectra were recorded with 500~$\mu$W probe power and 30~mW coupling power.}
 \label{fig:Spectra}
\end{figure}

After establishing main properties of the single-beam
NFE, a series of NFE spectra with presence of
the coupling beam was recorded. For these measurements the
probe-light frequency was scanned across the $D2$ line while the
coupling beam was tuned to the $5P_{3/2}\rightarrow 5D_{5/2}$
transition. NFE spectra measured with 500~$\mu$W probe
and 30~mW coupling beams and magnetic fields of $B=\pm 67$~mG are
shown in Fig.~\ref{fig:Spectra}.
Doppler-broadened signals were observed at
transitions associated with the resolved hyperfine structure of
the ground states of both isotopes of rubidium, $F=2\rightarrow
F'$ and $F=1\rightarrow F'$ for $^{87}$Rb and $F=3\rightarrow F'$
and $F=2\rightarrow F'$ for $^{85}$Rb, where $F'$ denote
unresolved excited-state hyperfine structure components.
Reversing direction of the magnetic field changed signs of the
observed dependence (Fig.~\ref{fig:Spectra}), which was consistent
with the change of the sign of rotation seen in
Fig.~\ref{fig:NFE}. Small overall background observed for both
spectra was related to imperfection of our polarimeter which
resulted in residual polarization rotation.

While hyperfine structure of the excited states was not resolved
with a single probe beam (Fig.~\ref{fig:SpectraWithoutCoupling}), application of a
counter-propagating coupling beam resulted in distinct
Doppler-free dips related to EIT (Fig.~\ref{fig:Spectra}). They
were nearly 100\% deep at the $F=2\rightarrow F'$ transition of
$^{87}$Rb and $F=3\rightarrow F'$ transition of $^{85}$Rb, but
weaker dips were also observed at the $F=2\rightarrow F'$
transition of $^{85}$Rb.

In a gas sample, the EIT-induced spectral features can be observed
only when the probe and the coupling beams simultaneously meet the
resonance conditions
\begin{equation}
 \begin{split}
  \omega_p+\vect{k_p}\cdot\vect{v}=\omega_p^0, \\
  \omega_c+\vect{k_c}\cdot\vect{v}=\omega_c^0,
  \label{eq:ResonanceConditions}
 \end{split}
\end{equation}
where $\vect{k_p}$ ($\vect{k_c}$) denotes the probe (coupling)
light wave vector, $\vect{v}$ is the atomic velocity, and
$\omega_p^0$ ($\omega_c^0$) is the central frequency of the
$5S_{1/2}\rightarrow 5P_{3/2}$ ($5P_{3/2}\rightarrow 5D_{5/2}$)
line. For counter-propagating beams
($\vect{k_p}\approx-\vect{k_c}$), observation of the dips
implies that the probe-light frequency fulfills the relation
\begin{equation}
 \omega_p=\omega_p^0+\frac{k_p}{k_c}(\omega_c^0-\omega_c).
 \label{eq:ProbeFrequency}
\end{equation}

With fixed coupling-light frequency and the probe scanned across
the transitions, the Doppler-free dips reveal hyperfine structure
of the top state of the ladder structure (the $5D_{5/2}$ state).
Inset to Fig.~\ref{fig:Spectra} presents the well resolved
hyperfine structure of the $5D_{5/2}$ state of $^{87}$Rb with
about 30 MHz spacings between neighboring components. At the same time the hyperfine structure of
the $^{85}$Rb $5D_{5/2}$ level remains unresolved within the
$F=3\rightarrow F'$ line, which is the consequence of a much
smaller hyperfine splitting of that state relative to the
$5D_{5/2}$ state of $^{87}$Rb.

Exact positions of the EIT dips and their assignments to specific
hyperfine components depend on particular tuning of the coupling
beam. For example, the NFE spectrum of $^{85}$Rb shown in
Fig.~\ref{fig:Spectra}, reveals different positions of the
Doppler-free EIT dips relative to the broad line centers. It can
be explained by taking into account different tunings of the
coupling beam with respect to the strongest transition within the
Doppler-broaden line. In particular, the spectrum shown in
Fig.~\ref{fig:Spectra} was taken with the coupling light tuned
close to the center of the $F'=4\rightarrow F''=5$ transition
which, according to Eq.~(\ref{eq:ProbeFrequency}), implies probe light to be tuned to  the $F=3\rightarrow F'=4$ transition for efficient EIT generation.
Since this transition determines the position of the
Doppler-broaden line, very deep dip appeared nearly at the center
of broad line $F=3\rightarrow F'$. However, the same coupling
frequency is detuned by 121~MHz from $F'=3\rightarrow F''=4$
transition. Since the strongest component of the $F=2\rightarrow
F'$ line is associated with the $F'=3$ state, the dip was shifted
with respect to the line center.

\begin{figure}
    \includegraphics{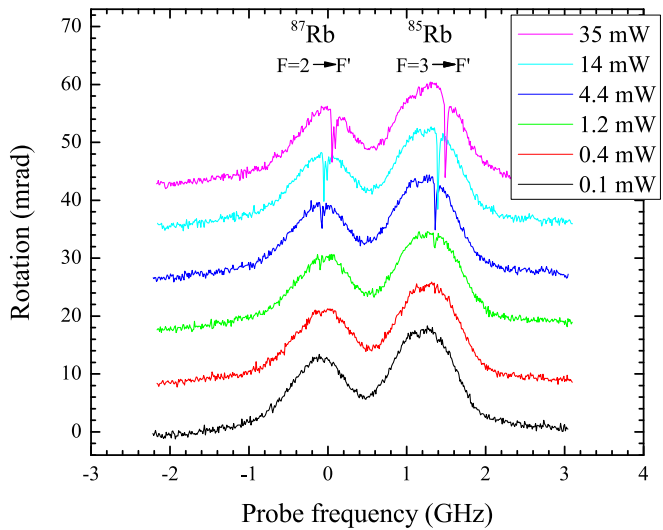}
    \caption{Dependence of the probe-beam NFE spectra on the coupling-light power
    for the $F=2\rightarrow F'$ ($^{87}$Rb) and $F=3\rightarrow  F'$ ($^{85}$Rb) components of the spectrum shown in Fig.~\ref{fig:Spectra}.
    For low power no Doppler-free dips are visible. For coupling-beam powers higher
    than 1~mW, Doppler-free resonances show-up due to EIT. For each
    spectrum the coupling-light frequency was detuned from the resonance which resulted in
    frequency shifts of the observed Doppler-free features. For clarity, the signals were
    vertically displaced. The spectra were recorded with $B=67$~mG.}
    \label{fig:SpectraPower}
\end{figure}

Having analyzed spectral characteristics of NFE with the coupling
beam, we investigated probe-polarization rotation versus the
coupling-beam power. Figure~\ref{fig:SpectraPower} presents such
dependence measured for two strongest Doppler-broadened NFE lines,
the $F=2\rightarrow F'$ transition of $^{87}$Rb and
$F=3\rightarrow F'$ of $^{85}$Rb. For coupling light power below
$P_c<0.1$~mW, no EIT was created and no Doppler-free features were
observed in the rotation spectra. However, for higher
coupling-beam powers, $P_c>0.1$~mW, EIT became effective and
resulted in appearance of Doppler-free dips within the spectrum.
In particular, similarly as in inset to Fig.~\ref{fig:Spectra},
well resolved features reflecting hyperfine structure of the
excited state $5D_{5/2}$ were seen at the $F=2\rightarrow F'$
transition of $^{87}$Rb. While the resonant dips reached nearly
100\% contrast, their power broadening was very small within a range of applied light intensities.

\begin{figure}
    \includegraphics{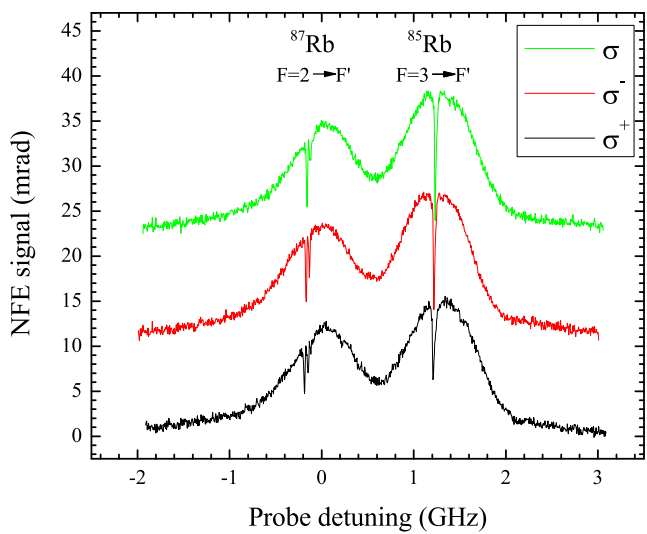}
    \caption{NFE spectra vs. probe-light frequency for different polarizations of the
    coupling beam. The spectra, in particular, the Doppler-free dips associated with EIT,
    did not depend on coupling-beam polarization. The measurements were performed with
    750~$\mu$W probe power, 14~mW coupling-beam power, and a magnetic field of 67~mG.
    For clarity the signals for $\sigma^+$ and $\sigma^-$ polarizations were vertically shifted.}
    \label{fig:SpectraPolarization}
\end{figure}

Figure~\ref{fig:SpectraPolarization} presents NFE
spectra measured with three different polarizations of the
coupling beam, two opposite circular and linear polarizations with
unchanged remaining experimental parameters. As shown, the spectra did not depend on
the coupling-beam polarization. This point differs significantly
from other measurements on coherent control of magneto-optical
rotation \cite{Liao1976,Liao1977,Pavone1997,Wielandy1998,Padney2008} and
indicates different nature of the rotation in our experiment than in previous studied cases. While in our experiment rotation was due to ground-state
coherences, in earlier works it was caused by excited-state
coherences and/or populations. This problem is addressed in more
detail in Sec.~\ref{sec:Theory}.

\begin{figure}
    \includegraphics{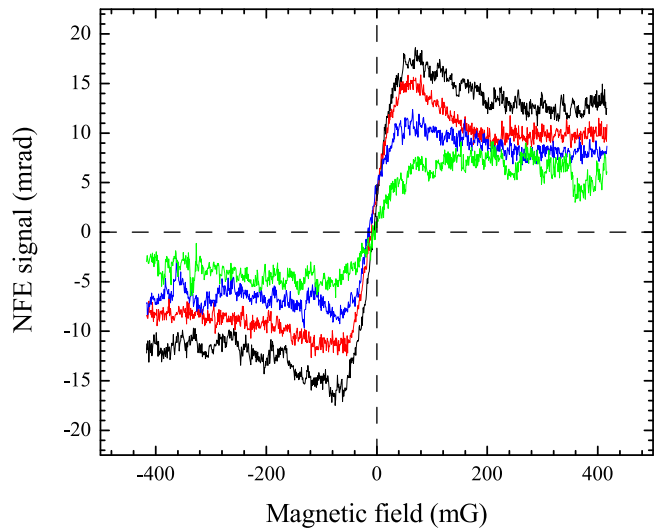}
    \caption{NFE signal vs. magnetic field for different tunings of the coupling beam.
    Polarization rotation was the strongest when probe and coupling beams interacted
    with atoms from different velocity classes, i.e., when conditions given by
    Eqs.~(\ref{eq:ResonanceConditions}) were not fulfilled (black curve).
    The amplitude of the signal deteriorated when the beams started to generate
    EIT between ground and excited states. The strongest suppression of the signal
    were observed for the probe and coupling beams fulfilling the resonance conditions
    and interacting with the same atoms (green curve). The signals were measured with
    $750$~$\mu$W-probe tuned to the center of the $F=3\rightarrow F'$ transition of
    $^{85}$Rb and $\sim 15$~mW power of the coupling beam.}
    \label{fig:NFEtuning}
\end{figure}

Experimental study were completed with the series of measurements
of NFE signals versus magnetic field for fixed probe frequency $\omega_p$ and
different coupling-beam detunings (Fig.~\ref{fig:NFEtuning}).
As seen, the amplitude of the signal deteriorates when the
coupling light is tuned to the two-photon resonance. Specifically,
it was almost three times smaller for the coupling beam in
resonance, i.e., when both beams fulfilled the conditions given by
Eqs.~(\ref{eq:ResonanceConditions}), than for far detuned coupling
light. This reduction of the NFE signal amplitude is
another manifestation of the competition between two processes of
coherences generation. Fulfilling the two-photon resonance
condition with appropriately strong coupling beam triggers EIT
which strongly modifies efficiency of ground-state coherence
generation and reduces the observed signal.

\section{Theoretical analysis\label{sec:Theory}}

Although rubidium energy levels involved in our experiment form a
fairly complex scheme [Fig.~\ref{fig:EnergyStructure}(a)], all salient features of our results can be interpreted in a much simpler \textit{inverted Y} model system
depicted in Fig.~\ref{fig:EnergyStructure}(b). It enables simple
analytical solutions of the density matrix and provides very good
physical insight into the problem. The model system consists of a
ground state with two Zeeman components corresponding to $m=\pm1$
(states $|\pm\rangle$) which are split in the external magnetic
field $B$ by 2$\omega_L$, and two excited states, one labelled as
$|0\rangle$ coupled with $|+\rangle$ and $|-\rangle$ by a linearly
polarized probe beam, and state $|e\rangle$ which is linked with
$|0\rangle$ by a strong coupling beam. Without the coupling beam
the model represents $\Lambda$ system which is a generic structure
not only for NFE, but also other CPT-like phenomena. With
just one of the ground-state sublevels, the model becomes a
familiar ladder EIT scheme. Simultaneous presence of two light
beams and a magnetic field allows interplay of both these effects.
Our analysis differs from earlier work on inverted Y scheme by
Joshi and Xiao \cite{Joshi2003} as it employs two, rather than
three laser beams in a combination of the ladder system and the
$\Lambda$ system which involves Zeeman sublevels in the lower
state. The latter system allows control of the atomic coherence by
a magnetic field.

We performed analysis of the experiment using the density-matrix
formalism in which the evolution of the density operator $\varrho$
is described by the Liouville equation
\begin{equation}
 \dot{\varrho}=-\frac{i}{\hbar}[H,\varrho]+\frac{1}{2}\{\Gamma,\varrho\}+\Lambda,
 \label{eq:LEq}
\end{equation}
where $[~,~]$ denotes commutator, $\{~,~\}$ anticommutator, $H$
the full Hamiltonian of the system, while $\Gamma$ and $\Lambda$
represent the relaxation and pumping operators, respectively,
responsible for retaining the thermal equilibrium populations of
the unperturbed system. In our analysis we took hamiltonian $H$ as
\begin{equation}
    H=H_0-\vect{\mu}\cdot\vect{B}-\vect{E}\cdot\vect{d},
    \label{eq:Hamiltonian}
\end{equation}
where $H_0$ is unperturbed Hamiltonian of the system, $\vect{d}$ denotes electric and $\vect{\mu}$ magnetic dipole moment operators, and $\vect{E}$ is the electric field of light taken in the form
\begin{equation}
    \vect{E}=(\vect{e_+}+\vect{e_-})E_p e^{i\omega_pt}+\vect{e_e}E_ee^{i\omega_ct},
    \label{eq:ElectricFieldOfLight}
\end{equation}
with $\vect{e_i}$ are respective polarization vectors.

Within the rotating wave approximation, the model system can be
described by the following set of equations for slowly-varying
envelopes $\sigma$ of the density matrix elements.

\begin{equation}
\begin{split}
\dot\rho_{\pm\pm}&=-i\Omega(\sigma_{\pm0}-\sigma_{0\pm})-\gamma(\rho_{\pm\pm}-N)+\frac{\Gamma_0}{2}\rho_{00},\\
\dot\rho_{00}&=i\Omega(\sigma_{+0}-\sigma_{0+}+\sigma_{-0}-\sigma_{0-})-i\beta(\sigma_{0e}-\sigma_{e0})\\
&\ \ \ -\Gamma_0\rho_{00}+\Gamma_e\rho_{ee},\\
\dot\rho_{ee}&=i\beta(\sigma_{0e}-\sigma_{e0})-\Gamma_e\rho_{ee},\\
\dot\sigma_{\pm0}&=-iA_\pm\sigma_{\pm0}+i\Omega(\rho_{00}-\rho_{\pm\pm}-\sigma_{\pm\mp})-i\beta\sigma_{\mp e},\\
\dot\sigma_{-+}&=iB\sigma_{-+}+i\Omega(\sigma_{0+}-\sigma_{-0}),\\
\dot\sigma_{0e}&=-iK\sigma_{0e}+i\Omega(\sigma_{-e}+\sigma_{+e})+i\beta(\rho_{ee}-\rho_{00}),\\
\dot\sigma_{\pm e}&=-iC_\pm\sigma_{\pm e}+i\Omega\sigma_{0e}-i\beta\sigma_{\pm 0}.\\
\end{split}
\label{eq:DensityMatrixRWA}
\end{equation}
with
\begin{equation}
\begin{split}
A_\pm&=\Delta\omega_0\pm\omega_L-i\gamma_{\pm 0},\\
K&=2\omega_L+i\gamma_{-+},\\
C_\pm&=\Delta\omega_0+\Delta\omega_e\pm\omega_L-i\gamma_{\pm e},\\
M&=\Delta\omega_e-i\gamma_{0e},\\
\end{split}
\label{eq:Substitution}
\end{equation}
where $\gamma_{ik}$ denote the relaxation rates of the
$\sigma_{ik}$ coherences ($i,k=±,0,e$), $\gamma$ is the relaxation
rate of the ground-state population, $\Gamma_0$ the spontaneous
emission rate of state $|0\rangle$, and $\Gamma_e$ of state
$|e\rangle$, $\Delta\omega_0$ and $\Delta\omega_e$ are detuning of
the probe and coupling beams from respective transitions at zero
magnetic field, $\omega_L$ is the Larmor frequency, $\Omega$ and
$\beta$ are the Rabi frequencies of the probe and coupling beams,
and $N$ denotes the unperturbed equilibrium populations of the
ground states. Equations~(\ref{eq:DensityMatrixRWA}) solved within
the steady-state approximation yield
\begin{equation}
\begin{split}
\sigma_{\pm0}&=-\frac{\Omega}{A_\pm}(N+\sigma_{\pm\mp})-\frac{\beta}{A_\pm}\sigma_{\pm e},\\
\sigma_{-+}&=\frac{\Omega}{K}(\sigma_{-0}-\sigma_{0+}),\\
\sigma_{0e}&=\frac{\Omega}{M}(\sigma_{-e}+\sigma_{+e}),\\
\sigma_{\pm e}&=-\frac{\beta}{C_\pm}\sigma_{\pm0},\\
\end{split}
\label{eq:DensityMatrixSSA}
\end{equation}
where the probe-beam Rabi frequency $\Omega$ was assumed to be
much smaller than the excite-state relaxation rate $\Gamma_0$,
($\Gamma_0\ll\Omega$) but stronger than the relaxation rate of the
ground state $\gamma$. It allows one to omit product
$\Omega\sigma_{0e}$ in Eqs.~(\ref{eq:DensityMatrixSSA})
\cite{Gea1995} and neglect optical pumping (within this
approximation the populations of the levels did not depend on
light intensity: $\rho_{\pm\pm}= N$ and
$\rho_{00}=\rho_{ee}=0$). These assumptions enable
presentation of the optical coherences $\sigma_{\pm 0}$, as well
as the $\sigma_{-+}$ and $\sigma_{\pm e}$ coherences responsible
for NFE and EIT in very simple analytical forms
\begin{equation}
\begin{split}
\sigma_{\pm 0}&=\frac{\Omega K C_\pm}{D_{\pm}}(A_\mp^*C_\mp^*+\beta^2)N,\\
\sigma_{\pm e}&=-\frac{\Omega \beta K}{D_{\pm}}(A_\mp^*C_\mp^*-\beta^2)N,\\
\sigma_{-+}&=\frac{\Omega^2}{D_-}\left[C_-C_+^*(A_+^*-A_-)-\beta^2(C_+^*-C_-)\right]N,\\
\end{split}
\label{eq:Coherences}
\end{equation}
where
\begin{equation}
\begin{split}
D_\pm=&C_\pm C_\mp^*[A_\pm A_\mp^*K-\Omega^2(A_\pm-A_\mp^*)]-\beta^2K\times\\
&\times(A_\pm C_\pm+A_\mp^*C_\mp^*)-\beta^2\Omega^2(C_\pm-C_\mp^*)+\beta^4K.\\
\end{split}
\label{eq:Ddefinition}
\end{equation}

It should be noted that despite these simplification
theoretical modeling based on Eqs.~(\ref{eq:Coherences}) are in
very good agreement with experimental data. Taking into account
optical pumping effects is quite straightforward wih our model but
does not change the overall character of the simulated results.

Equations~(\ref{eq:Coherences}) can be used for modelling various
signals. For instance, in the transmission measurements the
absorption coefficient for a single polarization component of the
probe is determined by Im$\sigma_{±0}$, in the magnetic dichroism
measurement the signal is given by Im$(\sigma_{+0}
-\sigma_{-0})$, and for detection of fluorescence its intensity
is given by $\text{Im}(\sigma_{+0} + \sigma_{-0})$. For the
rotation measurements, such as performed in this work, the Faraday
rotation angle is proportional to the difference of the real parts
of the optical coherences associated with the $\sigma^+$ and
$\sigma^-$ components of the probe field $\theta\propto
\text{Re}(\sigma_{+0}-\sigma_{-0})$.

\begin{figure}
 \includegraphics[width=\columnwidth]{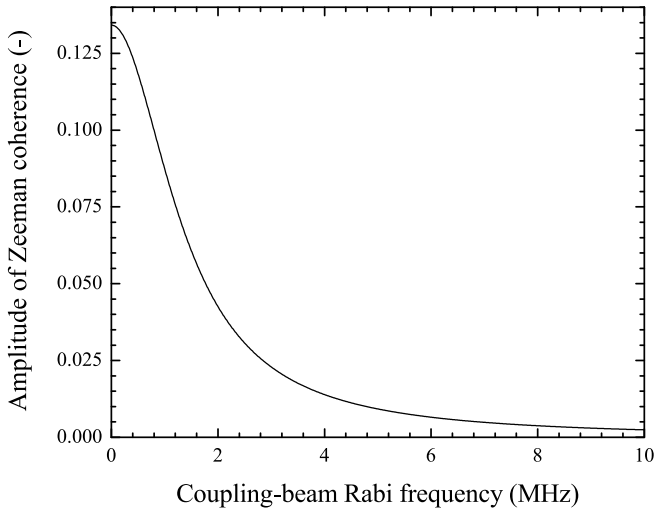}
 \caption{Amplitude of the real part of the Zeeman coherences $\sigma_{-+}$ as a
 function of the Rabi frequency of the coupling light $\beta$. Signals were simulated
 for both beams tuned to the resonance ($\Delta\omega_0=\Delta\omega_e=0$), probe-light
 Rabi frequency of 0.3~MHz, the Larmor frequency corresponding to the maximum of the NFE
 signal $\omega_L=0.05$~MHz, and the equilibrium populations $N=0.5$.}
  \label{fig:ZeemanCoherence}
\end{figure}

For $\beta=0$, Eqs.~(\ref{eq:Coherences}) allow one to reproduce
typical NFE signals such as depicted in
Fig.~\ref{fig:NFE}. When the coupling laser is on ($\beta\neq0$) but the
magnetic field is off ($B=0$), the formulae can be used
for description of EIT on the degenerate $|\pm\rangle$ -
$|0\rangle$ transitions. When all perturbations are acting
simultaneously ($\beta\neq0$, $B\neq0$), solutions~(\ref{eq:Coherences})
reproduce effects due to EIT and CPT, such as predicted by Joshi
and Xiao \cite{Joshi2003} for transmission measurements. For the
Faraday geometry, expressions~(\ref{eq:Coherences}) reflect
competition between NFE and EIT which not only bleaches the weak
probe absorption but also reduces nonlinear rotation of its
polarization, as shown in Figs.~\ref{fig:Spectra}-\ref{fig:NFEtuning}.
Figure~\ref{fig:ZeemanCoherence} shows how the amplitude of the real part of
the Zeeman coherence $\sigma_{-+}$ is reduced by increasing Rabi
frequency $\beta$ of the coupling laser. Since this is the Zeeman
coherence which is responsible for the nonlinear Faraday effect,
this dependence explains the reduction of the rotation signal
amplitude seen in the experiments with the coupling laser, i.e.
when $\beta\neq0$.

\begin{figure}
 \includegraphics[width=\columnwidth]{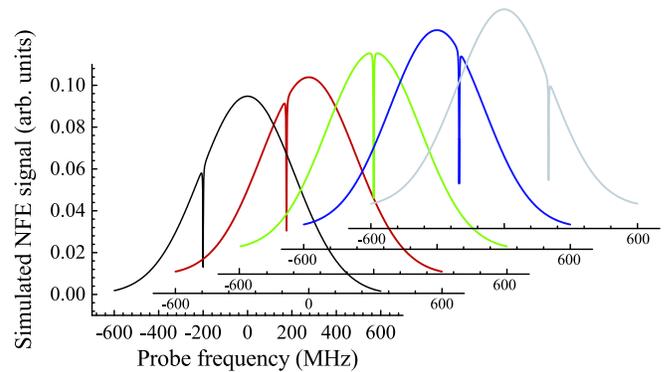}
 \caption{Theoretical NFE spectra calculated for $\Delta\omega_0=0$ and different detunings of the coupling
 laser: $\Delta\omega_e=200$~MHz (gray), 100~MHz (blue), 0~MHz (green), -100~MHz (red),
 and $-200$~MHz (black) from the bottom to the top curve, respectively. The signals were calculated for a Larmor frequency of 0.1~MHz, probe
 and coupling-light Rabi frequencies of $\Omega=0.3$~MHz and $\beta=3$~MHz, respectively.}
  \label{fig:Detunings}
\end{figure}

\begin{figure}
 \includegraphics[width=\columnwidth]{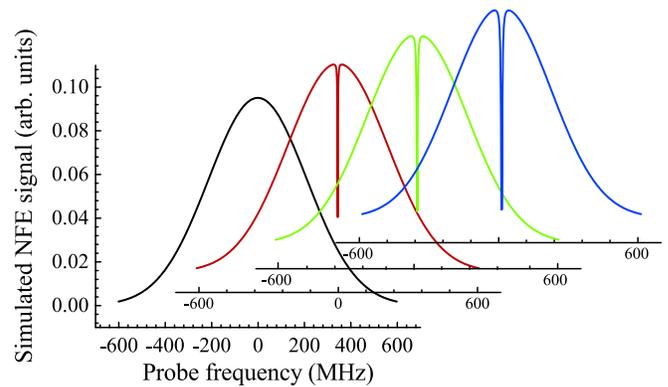}
 \caption{NFE spectra simulated for different Rabi frequencies of the
 coupling laser, $\beta =0, 3, 6$, and 9~MHz. The simulations were performed with Larmor
 frequency of 0.1 MHz, probe-light Rabi frequency of 0.3~MHz, and $\Delta\omega_0=\Delta\omega_e=0$.}
  \label{fig:SimulatedNFEPower}
\end{figure}

For the description of experiments with a vapor cell, the density
matrix needs to be velocity averaged with the detunings of the
counter-propagating probe and coupling laser beams modified to
account for the Doppler shifts: $\Delta\omega_0\rightarrow\Delta\omega_0 - kv_p\approx \Delta\omega_0 - kv$,
$\Delta\omega_e\rightarrow\Delta\omega_e+kv_e\approx \Delta\omega_e+kv$, where $v$ is the atomic speed
along the laser beam directions and $k$ the wavenumber. Figure~\ref{fig:Detunings} represents such
averaged signals simulated with the inverted Y model for different
coupled-beam detunings $\Delta\omega_e$ and scanned probe-laser detuning
$\Delta\omega_e$. The theoretical rotation spectra agree very well with
our observations. In particular, they exhibit Doppler-free dips
which appear at two-photon resonance, i.e. when $\Delta\omega_0+
\Delta\omega_e=0$ (Fig.~\ref{fig:Detunings}). If $\Delta\omega_e\neq0$, the dips are
shifted from $\Delta\omega_0=0$, as seen in the experimental spectra
(Fig.~\ref{fig:SpectraPower}).
These dips have bigger contrast than regular ladder EIT
resonances.  The contrast increases with $\beta$ which is
demonstrated in Fig.~\ref{fig:SimulatedNFEPower} and agrees with the observation
(Fig.~\ref{fig:SpectraPower}). At the same time, power broadening
of the EIT rotation dip is very small. These features may be
useful for laser frequency stabilization, although the dip
positions are not related directly to the atomic transitions but
depend on particular $\Delta\omega_e$.

\begin{figure}
 \includegraphics[width=\columnwidth]{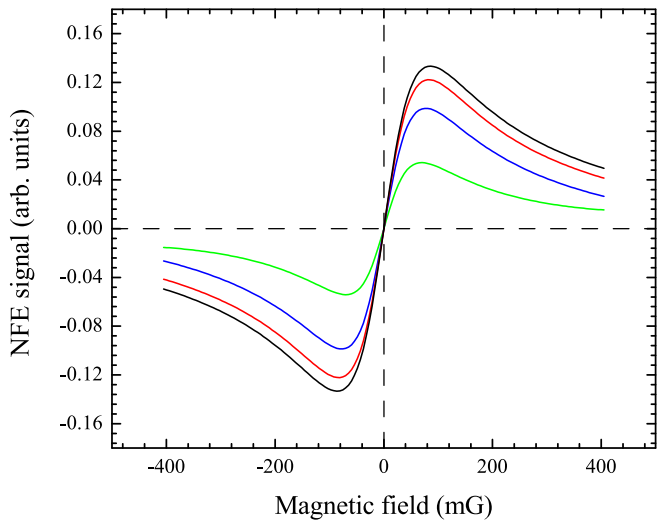}
 \caption{NFE signals vs. magnetic field simulated for probe and coupling-beam Rabi frequencies of 0.3~MHz and 3~MHz, respectively and different detunings of the coupling beam which correspond to those in Fig.~\ref{fig:NFEtuning}: $\Delta\omega_e=0.6$~MHz
 (green), 1.3~MHz (blue), 2.0~MHz (red), and 2.7~MHz (black) which manifest by different amplitudes of
 magneto-optical rotation.}
  \label{fig:SimulatedNFEtuning}
\end{figure}

The effect of quenching the Zeeman coherence by a coupling laser
illustrated in Fig.~\ref{fig:ZeemanCoherence} is responsible for
cancellation of the NFE signal. This effect is also very well
reproduced by our model. Figure~\ref{fig:SimulatedNFEtuning}
presents a sequence of the rotation signals calculated for
Doppler-averaged medium with fixed $B$ and $\omega_p$ and for
various values of $\Delta\omega_e$ which correspond to
those in Fig.~\ref{fig:NFEtuning}. The calculated signals
reproduce very well the observed behavior
(Fig.~\ref{fig:NFEtuning}).

In contrast to previous work on polarization control based on
light-induced optical anisotropy \cite{Wielandy1998, Padney2008},
the described changes of rotation  do not depend on the
coupling-beam polarization (Fig.~\ref{fig:SpectraPolarization}).
This difference is caused by the fact that in our study the
optical anisotropy resulting in the rotation of the probe beam
polarization reflects the ground-, rather than excited-state
coherence. The inverted Y model, very successful in explaining
most features observed in our experiment is quite insensitive to
the polarization of the coupling-beam  as this beam acts
symmetrically on both transitions involving the lower $\Lambda$ levels, hence it
does not create any anisotropy of the probe-beam propagation.

It is important to notice that while the real energy-level
structure of rubidium does possess many excited sublevels and many
possible excited-state coherences, these are the ground-state
coherences which dominate the probe-beam polarization rotation. This is because
the rates of creation of the ground state observables (coherences
and/or population redistribution) are by factor $\Gamma_0/\gamma$
bigger than the corresponding rates for the excited state. Thus,
if both coherences are allowed by the level structure, the
ground-state ones make by far stronger impact onto the interaction
dynamics (see, e.g., Ref. \cite{Gawlik1984}). However, the
contributions of the ground-state coherences to NFE signal have
the form of very narrow resonances, which reflects that they are
destroyed by fast precession in magnetic fields such that
$\omega_L>\gamma_{-+}$. For larger fields, such that $\gamma_{\-+}
\ll\omega_L\approx\Gamma_0$, the ground-state coherences became
irrelevant. Thus, for fields on the order of 10 G, the
excited-state observables become important and the rotation
depends on the coupling-beam polarization, as observed in
experiments with stronger magnetic fields
\cite{Wielandy1998,Padney2008}.

\section{Conclusions\label{sec:Conclusions}}

We have studied nonlinear Faraday effect under conditions of
electromagnetically induced transparency. It was experimentally discovered that the two coherent processes, NFE/CPT and EIT, compete which quenches
the magneto-optical rotation. A theoretical description based on a simple
inverted Y model reproduced all salient features of the
observations and revealed detailed physical mechanism behind the
observed behavior. The interplay between NFE and EIT allows one to
perform coherent control of the polarization state in a different
manner than in the previously proposed schemes. While the previous
approaches were relying on excited-state coherences, the described
method is based on the ground-state coherences which permits to
use much weaker light beams and magnetic fields to obtain the
desired control. The competition of NFE and EIT results in
Doppler-free spectral features which have higher contrast than the
regular EIT dips in a standard transmission measurements. In
principle, the enhanced contrast may be useful for precision
spectroscopy and/or laser-frequency stabilization.

\begin{acknowledgments}
This work was partly supported by the Polish Ministry of Science
grants NN505092033 and NN202074135.

\end{acknowledgments}

\end{document}